\documentclass{article}
\usepackage{amsmath}
\usepackage{amssymb}
\usepackage{graphicx}
\usepackage{hyperref}
\usepackage[inkscapelatex=false]{svg}
\hypersetup{
    colorlinks=true,
    linkcolor=blue,
    filecolor=black,
    urlcolor=black,
    citecolor=blue
}
\usepackage[utf8]{inputenc}
\usepackage[T1]{fontenc}
\usepackage{geometry}
\title{XASDB -- Design and Implementation of an Open-Access Spectral Database}

\author{
  \textbf{Denis Spasyuk} \\[1ex]
  Industry Services Group, Canadian Light Source \\[0.5ex]  
}
\date{}

\begin{document}

\maketitle
\begin{abstract}
The increasing volume and complexity of X-ray absorption spectroscopy (XAS) data generated at synchrotron facilities worldwide require robust infrastructure for data management, sharing, and analysis. This paper introduces the XAS Database (XASDB), a comprehensive web-based platform developed and hosted by the Canadian Light Source (CLS). The database houses more than 1000 reference spectra spanning 40 elements and 324 chemical compounds.

The platform employs a Node.js/MongoDB architecture designed to handle diverse data formats from multiple beamlines and synchrotron facilities. A key innovation is the XASproc JavaScript library, which enables browser-based XAS data processing including normalization, background subtraction, extended X-ray absorption fine structure (EXAFS) extraction, and preliminary analysis traditionally limited to desktop applications. The integrated XASVue spectral viewer provides installation-free data visualization and analysis with broad accessibility across devices and operating systems.

By offering standardized data output, comprehensive metadata, and integrated analytical capabilities, XASDB facilitates collaborative research and promotes FAIR (Findable, Accessible, Interoperable, and Reusable) data principles. The platform serves as a valuable resource for linear combination fitting (LCF) analysis, machine learning applications, and educational purposes. This initiative demonstrates the potential for web-centric approaches in XAS data analysis, accelerating advances in materials science, environmental research, chemistry, and biology.
\end{abstract}


\section{Introduction}
Advances in high-brilliance synchrotron light sources have revolutionized X-ray absorption spectroscopy (XAS), generating an increasingly large volume of complex and valuable data. To meet the growing need for efficient data management and accessibility, the Canadian Light Source (CLS) launched the development of the XAS Database (XASDB, https://xasdb.lightsource.ca) in 2018. Its purpose is to consolidate and make publicly available the extensive collection of XAS data produced by CLS staff and users, ensuring long-term preservation and broad accessibility. The database was publicly released in 2022 and received strong support, underscoring the importance of continuing the development of XASDB.

The demand for centralized XAS data repositories is not unique to CLS, with several other excellent initiatives existing worldwide. These include XASLIB (https://xaslib.xrayabsorption.org) \cite{XASLIB}, EELS (https://eelsdb.eu) \cite{EELS}, MDR XAFS DB (https://mdr.nims.go.jp/) \cite{MDR}, the SSHADE/FAME DB (https://wiki.sshade.eu) \cite{FAMESHADE}, the LISA XAS Database  (https://lisa.iom.cnr.it/xasdb) \cite{LISAXAS},  Experimental XAS database (http://xasdb.ihep.ac.cn) \cite{CHINAXASDB}, and RefXAS (http://xafsdb.ddns.net) \cite{REFXAS}. Together, these databases represent invaluable resources for the XAS community and emphasize the importance of FAIR (Findable, Accessible, Interoperable, and Reusable) data principles.

The need for efficient data sharing is especially critical for synchrotron facilities, which represent substantial investments in infrastructure, expertise, and energy. Collecting a single reference spectrum at the CLS is estimated to require roughly 40--100~kWh of electric power.\footnote{
    \parbox{\textwidth}{
        Estimated using:
        \[
        E_{\text{spectrum}} = \frac{E_{\text{total (kWh)}} \cdot t_{\text{spectrum (h)}}}{N_{\text{beamlines}} \cdot H_{\text{year}}},
        \]
        where $E_{\text{total (kWh)}}$ is the facility's annual energy, $t_{\text{spectrum (h)}}$ the spectrum collection time, $N_{\text{beamlines}}$ the number of beamlines, $H_{\text{year}}$ total hours in a year.
    }
} Maximizing the scientific return of these facilities therefore depends on making the data they generate readily accessible and interpretable by a broad research community. Robust data sharing minimizes redundant experiments, fosters interdisciplinary collaboration, and accelerates discovery, thereby amplifying the global impact of synchrotron science. 

The early stages of XASDB development employed the Python programming language and the Flask framework  for data processing and server-side interactions. Although Python provides powerful libraries for scientific computing and specifically for X-ray absorption fine structure (EXAFS) related analysis \cite{LARCH}, scaling the system into a robust and interactive Web application revealed limitations. Performance bottlenecks in real-time visualization, difficulties in maintaining two languages with different paradigms and tooling, and limited support for direct client-side interaction highlighted the need for a more web-centric architecture. To address these challenges, the project was re-engineered using Node.js \cite{NODEJS} for server-side operations and pure Vanilla JavaScript\cite{JAVASCRIPT} for the client-side interface. This paper presents the XASDB platform, outlining its design, implementation, and core features.

\section{XASDB Architecture and Implementation}

\subsection{System Overview}
XASDB is designed as a robust, highly responsive, and scalable web application. It employs a classic client-server architecture to ensure efficient data handling, rapid retrieval, and an interactive user experience. This architecture separates the data storage and processing logic (server-side) from the user interface and visualization (client-side). It allows for modular development, enhanced security, and improved performance. The server acts as the central repository and data preprocessing engine. The client-side provides a dynamic and responsive interface for users to access, visualize, and perform preliminary analysis on the XAS data.

The core of XASDB is a collection of reference spectra, each enriched with standardized metadata. To date, the database contains over 1000 spectra, spanning 40 elements and over 324 chemical compounds, with contributions from CLS staff \cite{MIRANDA}, CLS users \cite{VALERIE}, and other researchers \cite{MATTDATA}.

\subsection{Server-Side Implementation}
The server-side of the XASDB is built upon Node.js, chosen for its asynchronous, event-driven architecture, which is highly efficient for I/O-bound operations typical of web applications serving large datasets. Node.js enables fast data retrieval and processing, which is crucial to maintaining responsiveness even with numerous concurrent users.
For data storage, MongoDB \cite{MONGODB} was selected as the primary database. Its NoSQL, document-oriented nature provides significant flexibility, allowing for the storage of diverse XAS spectra with varying metadata structures without rigid schema constraints. This flexibility is particularly advantageous when incorporating data from different beamlines and synchrotrons, which often exhibit heterogeneous metadata and file formats. 

\begin{figure}[ht]
    \centering
    \includegraphics[width=0.75\linewidth]{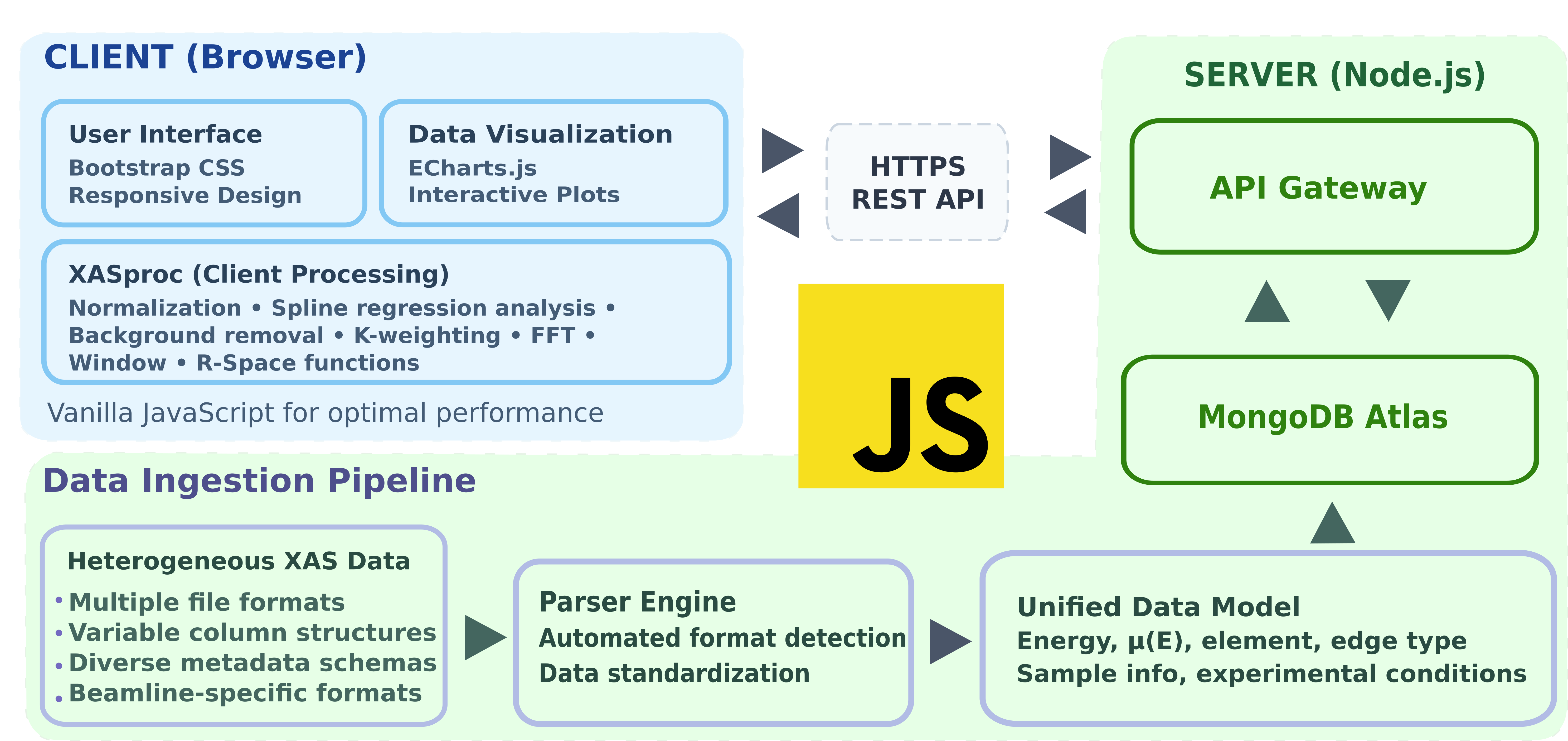}
    \caption{XASDB: Client-Server Architecture and Data Flow} 
    \label{fig: Diagram}
\end{figure}

A critical component on the server side is the \textbf{ }data parsing module. This module is responsible for ingesting raw XAS data files, which are typically organized in a hierarchical folder structure, and transforming them into a structured format suitable for efficient storage and retrieval in MongoDB. This module parses various common XAS data formats, extracts relevant spectroscopic data and metadata (e.g., energy, $\mu(E)$, element, edge, sample name, experimental conditions, beamline information) (Fig. \ref{fig: Diagram}).

Data heterogeneity posed a major challenge in database development due to the wide variation of file formats, naming conventions, column structures, and metadata standards across different beamlines and synchrotron facilities, highlighting the need for standardized, unified XAS formats such as XDI\cite{XDI} or CIF\cite{CIF}. To address this, the parsing module was designed with an extensible architecture, supporting multiple parsers capable of handling diverse input formats. During ingestion, these parsers normalize raw files by cleaning, restructuring, and standardizing both spectroscopic data and metadata. As a result, all records stored in MongoDB adhere to a consistent schema, enabling reliable querying, streamlined integration, and robust downstream analysis, regardless of the original data source. 

The database also supports user-driven search queries. Currently, the client-side user interface allows for searching by compound name, unique spectrum identifier (ID), and contributing author names, which are the fields currently indexed in MongoDB.

\subsection{Client-Side Implementation}
The client-side of the XASDB prioritizes a lightweight and high-performance user experience, achieved through the implementation of pure Vanilla JavaScript. Unlike larger client-side frameworks (e.g., React, Angular, and Vue), Vanilla JS provides direct control over the Document Object Model (DOM) and eliminates the overhead associated with framework-specific abstractions and dependencies. This choice contributes to faster page load times and a more fluid interactive experience, which is particularly important when dealing with real-time plotting and large data manipulation.

\begin{figure}[ht]
    \centering
    \includegraphics[width=0.75\linewidth]{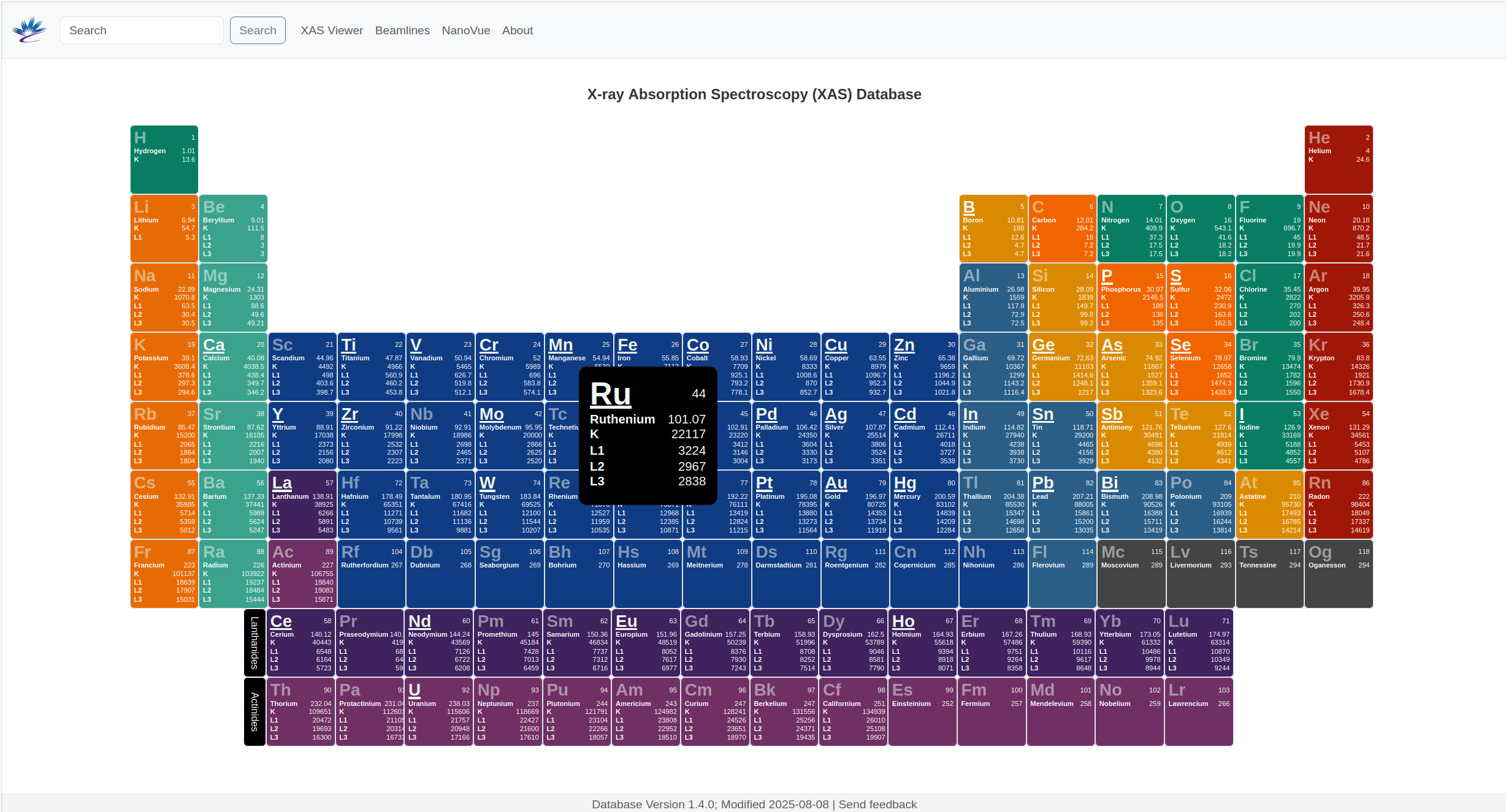}
    \caption{XASDB: Landing page}
    \label{fig:periodic}
\end{figure}

\begin{figure}[ht]
    \centering
    \includegraphics[width=0.75\linewidth]{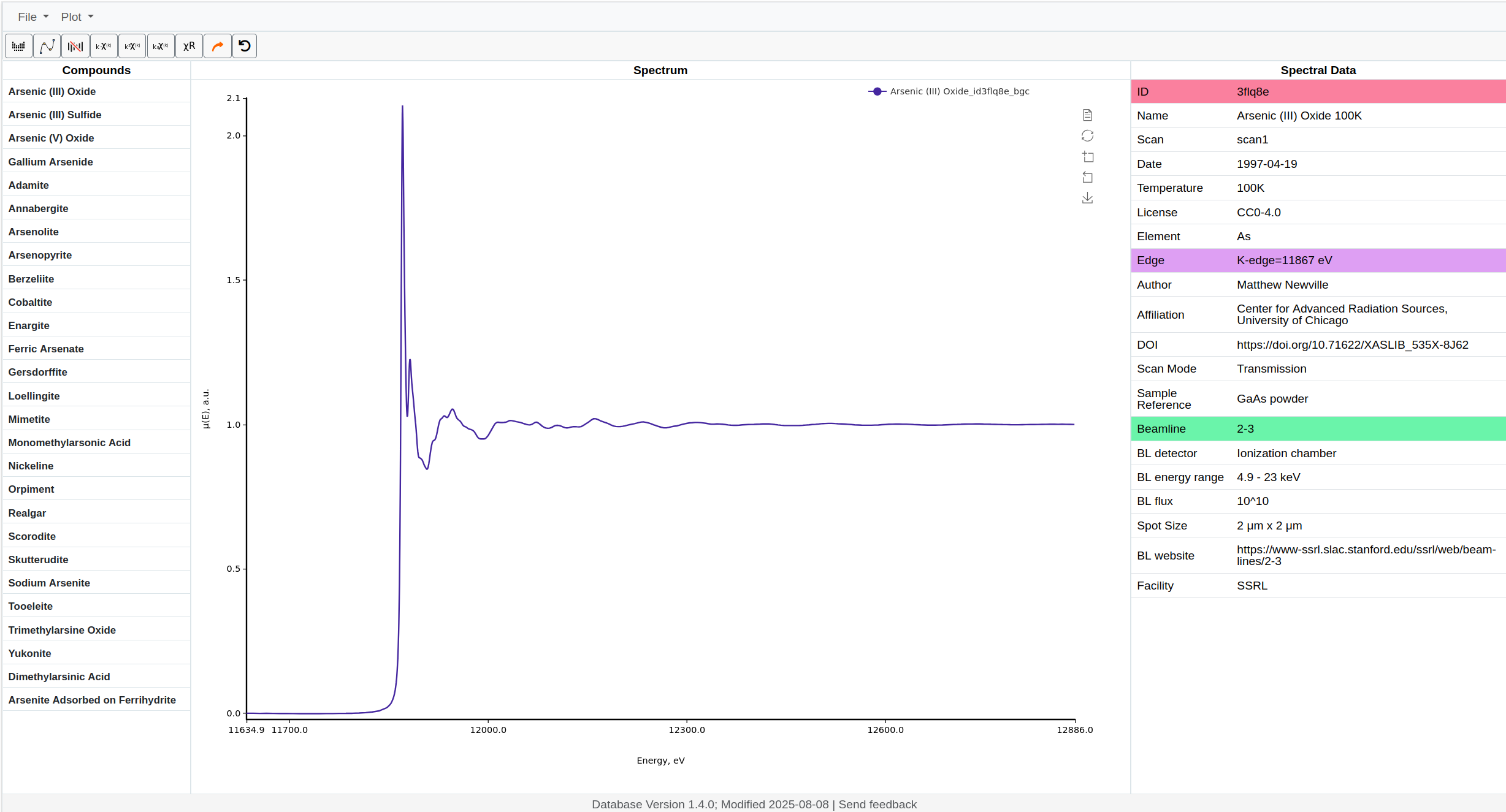}
    \caption{XASDB: Data Viewer}
    \label{fig:dataviewer}
\end{figure}

The user interface leverages Bootstrap CSS for its responsive design capabilities and pre-built UI components. Bootstrap ensures that the XASDB is accessible and visually consistent across various devices and screen sizes, from desktop workstations to mobile devices (Fig. \ref{fig:periodic} and \ref{fig:dataviewer}).

An important development for the client-side functionality is the XASproc module, a dedicated JavaScript library created specifically for the XASDB. This module encapsulates the mathematical routines and signal processing techniques required for XAS data manipulation. Unlike Python-based scientific ecosystems, JavaScript environments often lack specialized algorithmic tools for advanced data analysis. To address this gap, the necessary functions were implemented directly in XASproc, enabling accurate and automated XAS spectra processing within the browser. The library provides a comprehensive set of operations, including normalization, polynomial and B-spline fitting, background subtraction, derivative calculation, interpolation, extrapolation, \(k\)-weighting, Fourier transforms, \(R\)-space and \(Q\)-space calculations, and Savitzky--Golay smoothing. This allows users to perform preliminary data reduction without reliance on server-side computation.

The database supports two export formats for reference files: the original beamline software output and the standardized XDI format. Both formats can be exported through the \textit{File} menu in the client interface.

For visualization, the XASDB leverages the ECharts.js library \cite{ECHART}, which powers all interactive and dynamic graphical representations of XAS spectra.

\subsection{Data Licensing and Access}
All spectral data and associated metadata in the XASDB are made available under the Creative Commons Attribution 4.0 International License (CC BY 4.0). This licensing framework ensures that the database remains fully open-access while providing clear guidelines for data reuse and attribution. Users are free to download, distribute, remix, adapt, and build on the material in any medium or format, provided appropriate credit is given to the original data contributors and the XASDB platform. This permissive licensing approach aligns with the FAIR data principles and promotes maximum utility of the reference spectra for research, education, and commercial applications. The CC BY 4.0 license facilitates seamless integration of XASDB data into derivative works, machine learning training datasets, and collaborative research projects while maintaining proper attribution to the synchrotron facilities and researchers who contributed to the database.

\subsection{XASVue - Client-Side XAS Data Viewer and Analyzer}
\begin{figure}[ht]
    \centering
    \includegraphics[width=0.75\linewidth]{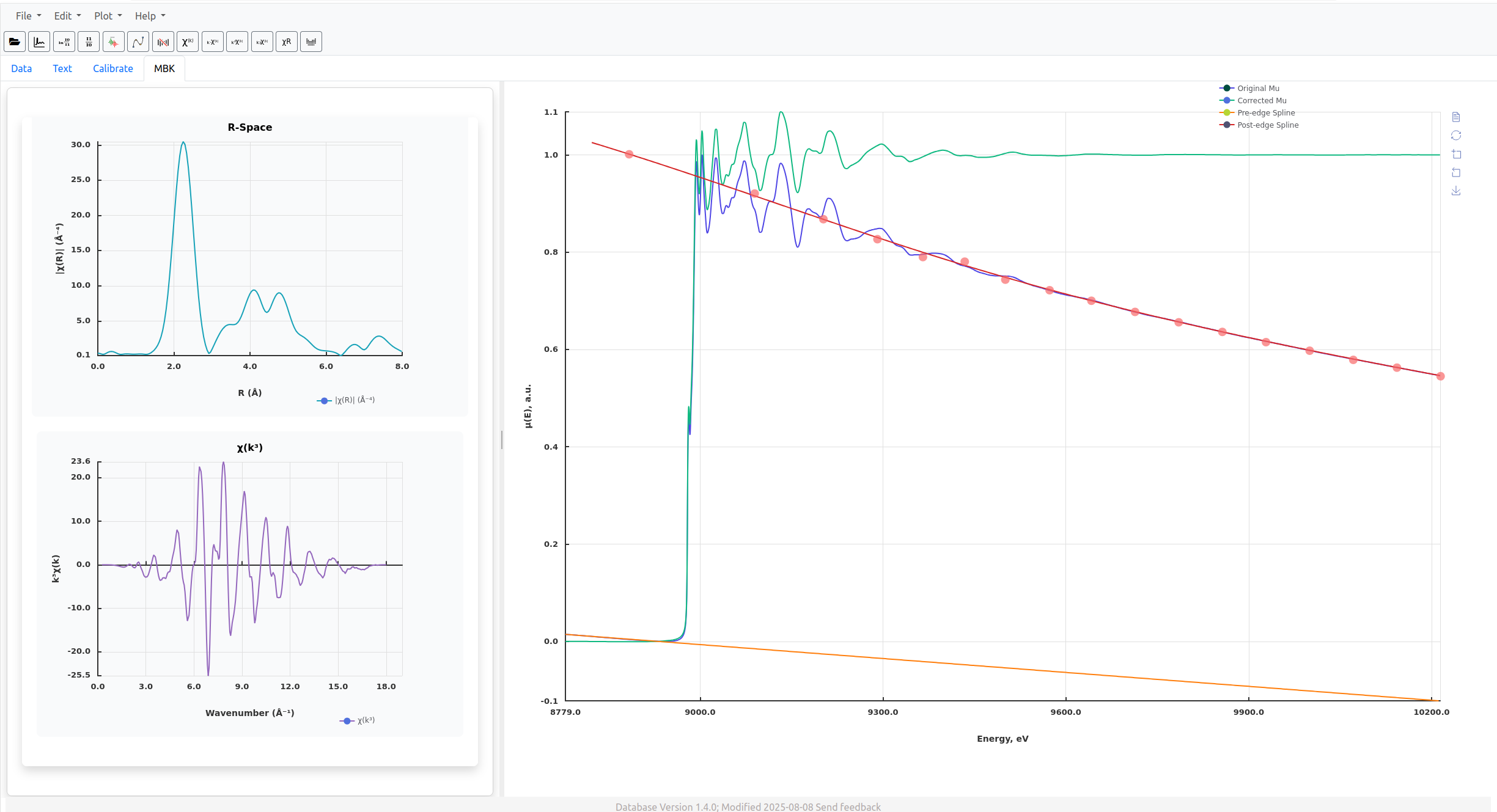}
    \caption{XASDB: XASVue - XAS file viewer}
    \label{fig:xasvue}
\end{figure}

XASVue (Fig. \ref{fig:xasvue}) is the integrated client-side application within the XASDB that serves as the primary graphical user interface for viewing and analyzing XAS data. It provides users with a simple and interactive environment, drawing parallels with established desktop XAS analysis software like Larch\cite{LARCH}, Athena\cite{ATHENA}, EXAFSPAK\cite{EXAFSPAK} and SIXPack\cite{SIXPACK}. A key advantage of XASVue is its readily accessible nature, requiring no installation on the user's local machine; it operates entirely within a user web browser, making it universally available to researchers regardless of their operating system or software environment. 

XASVue offers a suite of core functionalities designed to facilitate rapid data exploration and preliminary analysis. The integrated basic analysis capabilities, powered by the aforementioned XASproc module, allow users to perform common XAS data processing steps within the web interface. This immediate feedback loop helps users to quickly assess data quality, identify key features, and prepare spectra for more in-depth analysis using specialized software \cite{LARCH,ATHENA,EXAFSPAK, SIXPACK}. The seamless integration of the data viewer and analyzer with XASDB's comprehensive collection of XAS standards makes XASVue a powerful tool for fingerprinting, qualitative analysis, and education.

\section{XAS Data Processing Pipeline (XASproc)}

\subsection{Initial Data Preparation}
XAS data are typically collected in either transmission mode, fluorescence mode or Auger total electron yield (TEY). In transmission mode, the intensity of X-rays transmitted through the sample, $I_t$, is measured alongside the incident intensity, $I_0$. According to the Beer--Lambert law, the X-ray attenuation coefficient $\mu(E)$ is related to these intensities \cite{FUND} and the sample thickness $x$ via Eq.  \ref{eq:lac}. 
\begin{equation} \label{eq:lac}
\mu(E) = \frac{1}{x} \ln\left(\frac{I_0}{I_t}\right)
\end{equation}
In fluorescence mode, the signal is obtained by detecting photons or electrons emitted after core-level excitation. The measured fluorescence intensity $I_f$ is proportional to the product of the absorption coefficient $\mu(E)$ and the incident flux $I_0$. Under typical experimental conditions (thin sample, negligible self-absorption), this produces $\mu(E)$ according to Eq. \ref{eq:muflour} \cite{FUND}. 
\begin{equation}\label{eq:muflour}
\mu(E) \propto \frac{I_f}{I_0} 
\end{equation}
In XASDB, the software automatically identifies the appropriate columns for $I_0$, $I_t$, or $I_f$ depending on the acquisition mode, applies the relevant equations, and outputs $\mu(E)$. Users can override the automatic selection in XASVue, merge multiple detector channels, or process multiple datasets simultaneously before plotting. The processed data are then passed to the XASproc module for further analysis.

\subsection{Absorption Edge Determination ($E_0$)}
The XASproc module provides several methods for determining the absorption edge energy ($E_0$), which represents the energy at which the core electrons are excited to unoccupied states. The implementation includes several approaches, first-order derivative, smoothed first-order derivative, half-height method, and smoothed second-order derivative. Savitzky--Golay filter \cite{SG} is used to smooth the data before derivative calculation. 

By default the smoothed derivative function is used to find the absorption edge ($E_0$), which corresponds to the maximum in the first derivative of the $\mu(E)$ spectrum. This method is particularly effective for noisy spectra, as the smoothing step enhances the robustness of the derivative peak detection without compromising the resolution of the absorption edge. This is the default method used across the XAS database. 

\subsection{Background Subtraction and EXAFS Extraction}
Accurate background subtraction is a crucial preprocessing step in XAS for isolating the oscillatory EXAFS signal from the smooth atomic background. The XASproc module provides two complementary approaches for modeling and removing this background: a spline-based method and a weighted polynomial fitting method. 

\subsubsection{Spline-Based Background Subtraction}
The spline-based method fits a first- or second-order polynomial to the pre-edge and cubic smoothing splines to the post-edge regions of the absorption spectrum. The number of control points is calculated according to the modified Newville M. equation \cite{KNOTS} (Eq. \ref{eq:KNOTSM}).  The positions of the spline control points are equally spaced across the $k$-space and are not automatically refined. Users can adjust the positions of control points using an interactive graphical interface or optimize them using automatic refinement routines \cite{KNOTS}. To ensure the quality of the fit, a background quality score (BQS) (Eq. \ref{eq:bqs} ) is implemented. The BQS is a sum of the $k^3$-weighted mean offset (Eq. \ref{eq:chimean}), slope (Eq. \ref{eq:chislope}), symmetry (Eq. \ref{eq:chisymm}), and the square root of the spread of the $\chi k^3$ or variance ( Eq. \ref{eq:chivar} ). The obtained spline can then be applied for multiple scans of the same sample and used for further analysis. 
\begin{equation} \label{eq:bqs}
\text{BQS} = w_1 \cdot |\overline{\chi k^3}| 
+ w_2 \cdot \left| \frac{d (\chi k^3)}{dk} \right| 
+ w_3 \cdot \sqrt{\text{variance}}
+ w_4 \cdot \text{symmetry}
\end{equation}

\begin{equation} \label{eq:chimean}
\overline{\chi k^3} = \frac{1}{n} \sum_{i=1}^{n} \chi_i k_i^3
\end{equation}

\begin{equation} \label{eq:chislope}
\overline{k} = \frac{1}{n} \sum_{i=1}^{n} k_i, \quad
\text{slope} = \frac{\sum_{i=1}^{n} (k_i - \overline{k})(\chi_i k_i^3 - \overline{\chi k^3})}{\sum_{i=1}^{n} |k_i - \overline{k}|^p}
\end{equation}

\begin{equation} \label{eq:chisymm}
P = \sum_{\chi_i k_i^3 > 0} |\chi_i k_i^3|, \quad
N = \sum_{\chi_i k_i^3 < 0} |\chi_i k_i^3|, \quad
\text{symmetry} = 
\begin{cases} 
\dfrac{|P-N|}{P+N}, & P+N \neq 0 \\[2mm]
0, & P+N = 0
\end{cases}
\end{equation}

\begin{equation} \label{eq:chivar}
\text{variance} = \frac{1}{n} \sum_{i=1}^{n} (\chi_i k_i^3 - \overline{\chi k^3})^2
\end{equation}

\begin{equation} \label{eq:KNOTSM}
N_{knots} \approx 0.326 \cdot R_{bkg} \left( \sqrt{E_{max} - E_0} - \sqrt{E_{min} - E_0} \right)
\end{equation}

Where is:
\begin{itemize}
\item $\chi_i$: EXAFS oscillations at index $i$
\item $k_i$: photoelectron wavevector at index $i$
\item $n$: number of data points
\item $w_1, w_2, w_3, w_4$: component weights, default 1 for all
\item 0.326 a constant approximating $  \frac{2}{\pi} \sqrt{\frac{2m_e}{\hbar^2}}$ 
\item \(m_e\) is the electron rest mass.
\item \(\hbar\) is the reduced Planck constant.
\end{itemize}

\subsubsection{Weighted Polynomial-based Background Subtraction}
The polynomial-based method provides a flexible, automated framework for modeling pre-edge and post-edge backgrounds in XAFS data. Although not necessarily the most accurate approach, it excels at quick (500x faster than AUTOBK \cite{KNOTS}), robust processing of diverse datasets, including challenging high-noise spectra, with minimal user intervention.

The algorithm implements an adaptive approach that combines traditional polynomial regression with strategic data weighting to improve fitting accuracy in the EXAFS region. 
Polynomial order adapts automatically on the basis of point density, high energy coverage, and energy range. The algorithm selects third-order polynomials for high-quality, long-range datasets and defaults to the base order two for limited datasets, ensuring optimal balance between flexibility and stability across varying experimental conditions. To address poor fitting at the tail of the EXAFS region, it applies enhanced weighting strategies to points within the high-energy region defined as Eq. \ref{eq:weighted}.
\begin{equation} \label{eq:weighted}
E \geq E_{\text{threshold}} = E_{\text{max}} - (E_{\text{max}} - E_0) \times f_{\text{range}}
\end{equation}
 $f_{\text{range}}$ is the high-energy range fraction (typically 0.6), and $E_{\text{max}}$ is the maximum energy in the spectrum. Additional data points are generated through linear interpolation within the high-energy region to increase point density by a factor of 5. This approach only affects the polynomial fit, and not the actual data; it improves polynomial fitting at the spectrum terminus and minimizes polynomial end effects.

\subsubsection{Background Correction and Normalization}
The background correction procedure begins with the pre-edge region, which is fit with a linear function, $S_{\text{pre}}(E)$. Subtracting this function sets the pre-edge baseline at zero. The edge step height, $\Delta\mu$, is then determined as the difference between the extrapolated pre-edge and post-edge polynomial functions at the edge energy $E_0$.
The residual curvature in the post-edge region is accounted for by fitting a weighted polynomial function or spline functions $S_{\text{post}}(E)$, above the edge. Subtracting this function and normalizing by the edge step yields the corrected and flattened linear attenuation coefficient (Eq. \ref{eq:bgcorr}).
\begin{equation} \label{eq:bgcorr} \mu_{\text{corrected}}(E) = \begin{cases} \dfrac{\mu(E) - S_{\text{pre}}(E)}{\Delta\mu}, & E \leq E_0, \\[0.75em] \dfrac{\mu(E) - S_{\text{post}}(E) + \Delta\mu}{\Delta\mu}, & E > E_0. \end{cases} \end{equation}

This procedure establishes a baseline at zero before the edge, normalizes the absorption step to unity, and flattens the post-edge region. The resulting spectrum enables consistent comparison of  XANES features and reliable extraction of EXAFS oscillations.

\subsection{EXAFS $\chi(k)$ Calculation and $k$-Weighting}
The XASproc module also contains functions to extract the EXAFS oscillations, $\chi(k)$, from the background-subtracted $\mu(E)$ spectrum and calculate different k-weights. This helps to assess spectrum quality, noise levels, and potential glitches in the reference spectra. The EXAFS function is calculated \cite{FUND} using the standard EXAFS function (Eq. \ref{eq:chi}). The energy scale $E$ is then converted to the photoelectron wavevector $k$ \cite{FUND} using the relation (Eq. \ref{eq:key}):

\begin{equation} \label{eq:chi}
\chi(E) = \frac{\mu(E) - \mu_0(E)}{\Delta \mu_0} 
\end{equation}

\begin{equation} \label{eq:key}
k = \sqrt{\frac{2m_e(E - E_0)}{\hbar^2}}
\end{equation}

Following the extraction of $\chi(k)$, XASproc applies $k^n$ weighting to emphasize different regions of the EXAFS signal, where $n = 1, 2, 3$. The $k$-weighted EXAFS functions are defined as (Eq. \ref{eq:chik}) with $n=0$ corresponding to the unweighted $\chi(k)$. The resulting $k$-weighted EXAFS values can be exported from XASVue \textit{File} menu for further analysis.
\begin{equation}\label{eq:chik}
\chi_k^{(n)}(k) = k^n \chi(k), \quad n = 0, 1, 2, 3
\end{equation}

\subsection{Forward Fourier Transform with Riemann Approximation}
The XASproc also implements a robust method to transform EXAFS data from k-space ($\chi$(k)) to R-space ($\chi$(R)) while performing background subtraction using a Fourier filtering approach. The forward Fourier transform from k-space to R-space is computed using the standard EXAFS convention \cite{FUNDMATT} via the Riemann sum (Eq. \ref{eq:chir}).
\begin{equation} \label{eq:chir}
\chi(R) \approx \sqrt{\frac{2}{\pi}} \sum_{i=0}^{N-1} \chi_k(k_i) W(k_i) e^{-2 i k_i R} \Delta k
\end{equation}

For accurate numerical integration, the experimental data is first interpolated onto a uniform k-grid. The step size for this grid, $\Delta$k, is dynamically calculated based on the user-specified maximum R-space distance, Rmax (default 8\AA ). This ensures that the data are adequately sampled to represent the desired range. Similarly, the R-space step size, $\Delta$R, is set according to the total k-range of the data, (kmax--kmin), to maintain the highest possible R-space resolution. An 8x oversampling factor is applied to both $\Delta$k and $\Delta$R to produce a smooth, artifact-free spectrum and ensure optimal numerical precision.
In this implementation, no phase correction is applied, which maintains the natural phase relationships in the EXAFS oscillations. This results in slightly shorter atomic distances, by about 0.3 - 0.4 \AA.

To isolate physically significant EXAFS oscillations, a background cutoff Rbkg (default = 1.0 \AA) is applied to remove low-R artifacts and residual background. This filtered $\chi$(R) is then transformed back to k-space using the inverse Fourier transform \cite{FUNDMATT} (Eq. \ref{eq:inversechir}).

\begin{equation}
\label{eq:inversechir}
\chi_k^{\text{bkg}}(k) \approx \text{Re}\left[ \sqrt{\frac{2}{\pi}} \sum_{i=0}^{M-1} \chi_{\text{bkg}}(R_i) e^{i(2kR_i)} \Delta R \right]
\end{equation}
  Finally, the background-subtracted k-space signal is obtained (Eq. \ref{eq:chirbgsub}).
\begin{equation} \label{eq:chirbgsub}
\chi_k^\text{filtered}(k) = \chi_k(k) - \chi_k^\text{bkg}(k)
\end{equation}
   A second forward Fourier transform of $\chi_k^\text{filtered}(k)$ yields the background-subtracted $R$-space spectrum, $\chi(R)_\text{filtered}$. 

\subsection{Windowing Functions for EXAFS Data Processing}
In EXAFS data analysis, windowing functions are essential for isolating the desired k-space region and minimizing Fourier transform artifacts. The unified windowing function applies a window $W(k)$ to the EXAFS signal $\chi(k)$ according to:
\begin{equation}\label{eq:exafswin}
\chi_w(k) = \chi(k) \cdot W(k)
\end{equation}
where $\chi_w(k)$ is the windowed EXAFS signal and $W(k)$ is the window function.
For k-space windowing, the window function is defined over a specific k-range $[k_{\min}, k_{\max}]$ with tapering regions of width $\Delta k$ at both edges:
\begin{equation}\label{eq:wingen} W(k) = \begin{cases}
0 & \text{if } k < k_{\min} \\
w_{\text{taper}}\left(\frac{k - k_{\min}}{\Delta k}\right) & \text{if } k_{\min} \leq k < k_{\min} + \Delta k \\
1 & \text{if } k_{\min} + \Delta k \leq k \leq k_{\max} - \Delta k \\
w_{\text{taper}}\left(\frac{k_{\max} - k}{\Delta k}\right) & \text{if } k_{\max} - \Delta k < k \leq k_{\max} \\
0 & \text{if } k > k_{\max}
\end{cases}
\end{equation}
where $w_{\text{taper}}(n)$ is the tapering function with $n \in [0,1]$.

The XASproc module implements two distinct windowing methods: the Hanning window, optimized for general EXAFS applications, and the Kaiser window.

\subsubsection{Hanning Window (K-Space)}
The Hanning window \cite{HANNING} uses a cosine-based tapering function (Eq. \ref{eq:hanning}):
\begin{equation}\label{eq:hanning}
w_{\text{taper}}(n) = \frac{1}{2}\left(1 - \cos(\pi n)\right)
\end{equation}
The Hanning window is particularly well-suited for EXAFS applications due to its optimal balance between spectral resolution and artifact suppression. Its smooth cosine-based tapering provides excellent suppression of Fourier transform ringing while maintaining good frequency resolution, making it the most versatile choice for routine XAS data processing. This function is adopted as the default windowing function in major software packages \cite{LARCH,ATHENA} and is default in XASDB.

\subsubsection{Kaiser Window (K-Space)}
The Kaiser window \cite{KAISER} uses a modified Bessel \cite{BESSEL} function of the first kind for tapering (Eq. \ref{eq:kasert}):
\begin{equation}\label{eq:kasert}
w_{\text{taper}}(n) = \frac{I_0\left(\alpha\sqrt{1-(2n-1)^2}\right)}{I_0(\alpha)}
\end{equation}
where $I_0(x)$ is the modified Bessel function of the first kind of order zero (Eq. \ref{eq:kaseri}):
\begin{equation}\label{eq:kaseri}
I_0(x) = \sum_{m=0}^{\infty} \frac{1}{(m!)^2}\left(\frac{x}{2}\right)^{2m}
\end{equation}
and $\alpha$ is the Kaiser window parameter controlling the trade-off between main lobe width and side lobe suppression.

\section{Comparison of XASVue with Athena}
\begin{figure}[ht]
    \centering
    \includegraphics[width=0.75\linewidth]{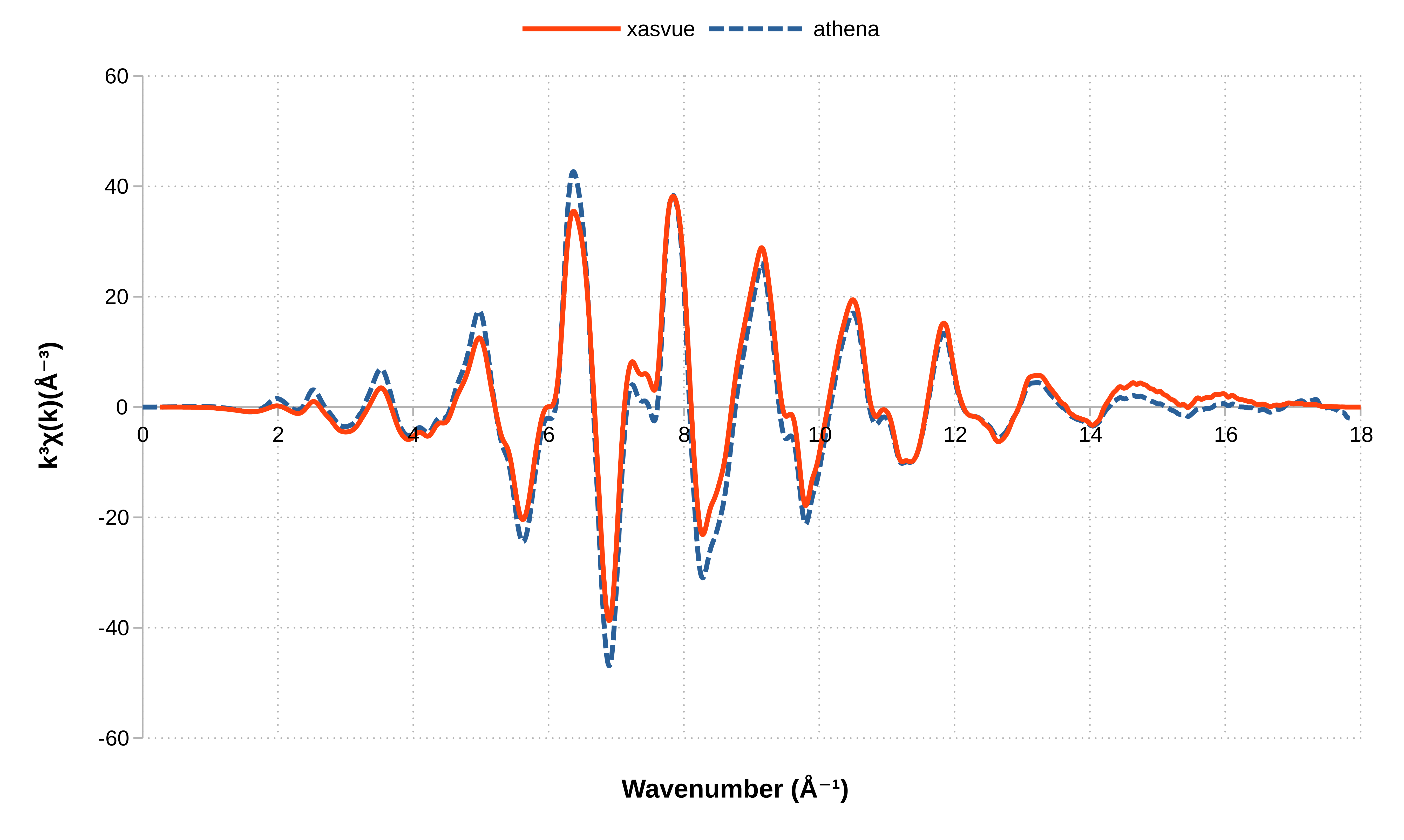}
    \caption{$k^3$-weighted EXAFS: Athena vs XASVue}
    \label{fig:k3space}
\end{figure}

In order to assess the performance of XASVue relative to the widely used
Athena package \cite{ATHENA}, we carried out a direct comparison of the processed extended EXAFS spectra in both $k^3$ weighted $k$-space and $R$-space. For this purpose, copper foil \cite{DESMAU} and the same processing parameters ($E_0$, d$k$, dR, etc.) were used. As shown in Fig.~\ref{fig:k3space}, the $k^3$-weighted EXAFS oscillations obtained from XASVue are very similar to those generated with Athena. 

\begin{figure}[ht]
    \centering
    \includegraphics[width=0.75\linewidth]{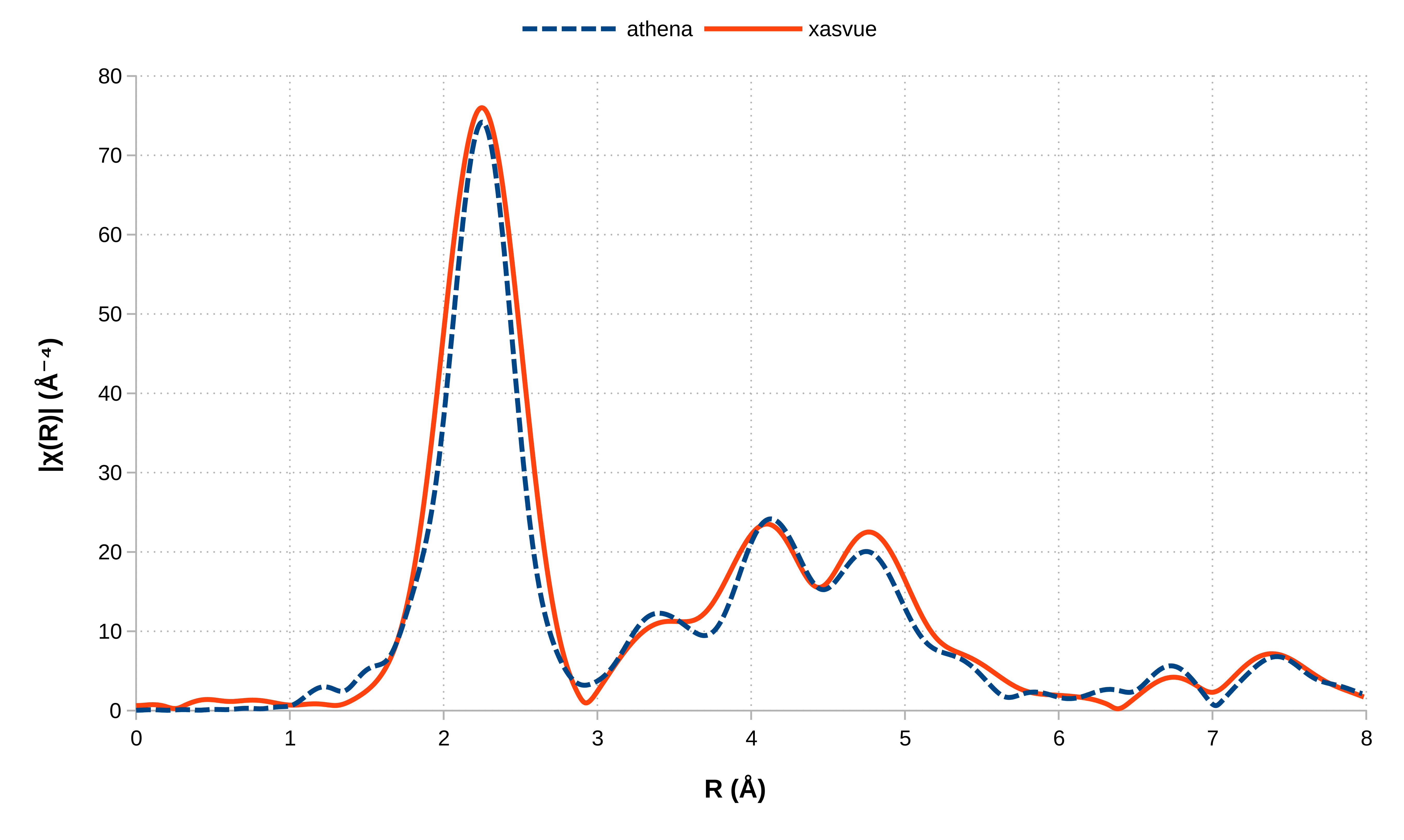}
    \caption{R-space: Athena vs XASVue}
    \label{fig:rspace}
\end{figure}

Similarly, the $R$-space (Fig.~\ref{fig:rspace}) reveal only marginal differences between the two programs, with the peak positions remaining the same. This spectral comparison shows that the mathematical treatment of EXAFS data within XASVue is consistent with the established standards of analysis.

\section*{Conclusions}
The XASDB platform successfully integrates a robust open-access spectral database with high-performance, browser-based analysis tools (XASproc, XASVue). By providing standardized data, comprehensive metadata, and essential processing capabilities, XASDB directly addresses the growing need for efficient data management in synchrotron science. Performance improvements realized through the Node.js and Vanilla JavaScript architecture have created a powerful and responsive user experience that can scale to meet the growing demands of synchrotron facilities.

By making XAS data freely available under the CC BY 4.0 license, XASDB promotes the principles of FAIR data. The standardized format and metadata make XASDB a valuable resource for training advanced analytical tools, including machine learning models for automated phase identification and oxidation state determination. We are hopeful that XASDB will become an invaluable tool for researchers, educators, and students to accelerate scientific advancements in geology, materials science, chemistry, and biology. Looking ahead, we envision XASDB evolving into a community-driven platform where users can securely contribute their own reference spectra, further enriching the database and ensuring its long-term sustainability and relevance.

\section*{Acknowledgements}
The author is deeply grateful to the Canadian Light Source and its staff -- Miranda Vu, Mehrnaz Mikhchian, Peter Blanchard, Joel Reid, Jeff Warner, Kurt Nienaber, Danielle Veikle, Mohsen Shakouri, Ning Chen, Erika Bergen, Amani Ebrahim, Morgane Desmau, and Roman Chernikov -- for their contributions of data and invaluable assistance in assembling and curating the XAS reference dataset.

The author also wishes to thank the many users and independent contributors who provided data for this project, including Matthew Newville, Bruce Ravel, Valerie A. Schoepfer, Heather E. Jamieson, Matthew B. J. Lindsay, Yuanming Pan, Jinru Lin, Daniel S. Alessi, and Andrew P. Grosvenor.

Additionally, the author thanks the broader CLS community, and specifically Michel Fodje, Stuart Read, and Toby Bond, for their constructive feedback. Any remaining errors or omissions are the author’s alone.

\bibliography{references}

\end{document}